\documentclass[twocolumn,showpacs,amsmath,amssymb,prc,superscriptaddress,floatfix]{revtex4}
\usepackage{amsmath,amssymb}
\usepackage{graphicx}
\usepackage{graphics}
\usepackage{epsfig}

\newcommand{\dd}{\mbox{$\textrm{d}$}}
\begin{document}
\title{Phenomenological study of the $\boldsymbol{pp\to\pi^+ pn}$ reaction}

\author{G.~F\"{a}ldt}\email[E-mail: ]{goran.faldt@physics.uu.se}
\affiliation{Department of Physics and Astronomy, Uppsala University, Box 516, 751 20 Uppsala, Sweden}
\author{C.~Wilkin}\email[E-mail: ]{c.wilkin@ucl.ac.uk}
\affiliation{Physics and Astronomy Department, UCL, Gower Street, London WC1E
6BT, United Kingdom}

\date{\today}
\begin{abstract}
Fully constrained bubble chamber data on the $pp\to \pi^+pn$ and $pp\to
\pi^+d$ reactions are used to investigate the ratio of the counting rates for
the two processes as function of the $pn$ excitation energy $Q$. Though it is
important to include effects associated with the $p$-wave nature of pion
production, the data are insufficient to establish unambiguously the
dependence on $Q$. The angular distributions show the presence of higher
partial waves which seem to be anomalously large at small $Q$. The dispersion
relation method to determine scattering lengths is extended to encompass
cases where, as for the $pp\to \pi^+pn$ reaction, there is a bound state and,
in a test example, it is shown that the values deduced for the low energy
neutron-proton scattering parameters are significantly influenced by the pion
$p$-wave behavior.
\end{abstract}
 %$(p, \pi)$ reactions
\maketitle

\section{Introduction}
\label{sec1}

The most complete measurements of the $pp\to\pi^+ pn$ differential cross
section were carried out using the PNPI bubble chamber exposed to proton
beams with kinetic energies between 900 and
1000~MeV~\cite{ERM2014,ERM2011,ERM2017}. The four-vectors of the final
particles for all the events at the three energies studied are available on
the Bonn-Gatchina WEB site~\cite{BG2017}. Although it was shown that much of
the data could be described through the excitation of the $\Delta(1232)$
isobar through pion exchange~\cite{ERM2014,ERM2011,ERM2017}, it is of
interest to see what features could be explained using less prescriptive
model approaches.

The final state interaction (FSI) theorem~\cite{FAL1997} links the production
of $S$-wave spin-triplet $pn$ pairs in the $pp\to\pi^+ pn$ reaction to the
cross section for $pp\to\pi^+ d$. The failure of the theorem to describe the
bubble chamber data was ascribed to the production of higher partial waves in
the recoiling proton-neutron system at even low $pn$ excitation energy
$Q$~\cite{FAL2017}. By studying the ratio of the $pp\to\pi^+ pn$ and
$pp\to\pi^+ d$ cross sections as a function of $Q$, as well as the angular
distribution of the produced $pn$ pairs, it is shown in Sec.~\ref{sec2} that
the behavior for $Q<20$~MeV is anomalous, possibly due to the strong tensor
force that couples the $S$ and $D$-waves. Though the $p$-wave nature of pion
production has to be taken into account when evaluating the predictions, this
does not affect the basic conclusions.

Though it is hard with the present data to isolate the contribution from
higher partial waves just on the basis of the FSI theorem, we turn in
Sec.~\ref{sec3} to the question of whether $pp\to\pi^+ pn$ could in principle
be used to investigate the properties of the low energy $pn$ system. Though
data on $pp\to K^+\Lambda p$ have been directly fitted to determine the
$\Lambda p$ scattering length~\cite{BUD2010}, an alternative approach has
been advocated that uses a dispersion relation in an approximate treatment
that only requires data over a limited range in
$Q$~\cite{GAS2004,GAS2005,HAU2017}. In the derivation of this formalism it is
assumed that there is no true bound state pole and in Sec.~\ref{sec3} we
generalize this method to treat the $pp\to\pi^+ pn$ reaction, where the final
$pn$ pair could combine to produce a true bound state, namely the deuteron.
By considering the predictions of a simplified model, the predicted $pn$
parameters are studied as functions of the cut-off in the dispersion relation
description with and without considering the $p$-wave nature of pion
production. In analogy to our analysis of the $pp\to K^+\Lambda p$
reaction~\cite{FAL2016}, it is shown that it is the position of the nearby
pole in the $pn$ channel that remains completely stable and the scattering
length itself is much more model-dependent.

Our conclusions are to be found in Sec.~\ref{sec4}.

\section{Comparison of the $\boldsymbol{pp\to\pi^+ pn}$ and $\boldsymbol{pp\to \pi^+d}$ production rates}
\label{sec2}

The final state interaction theorem relates the normalizations of the wave
functions for $S$-wave bound and scattering states~\cite{FAL1997}. This has
been exploited to predict the double-differential center-of-mass (cm) cross
section for the $S$-wave spin-triplet component in $pp\to\pi^+pn$ in terms of
the cross section for $pp\to\pi^+d$~\cite{BOU1996}:
\begin{eqnarray}\nonumber
\frac{\dd^2\sigma}{\dd\Omega\,\dd{x}}
(pp\to\pi^+\left\{pn\right\}_t) &=&\\
&&\hspace{-2.5cm}F(x)\,\frac{p_{\pi}(x)}{p_{\pi}(-1)}
\frac{\sqrt{x}}{2\pi(x+1)}\,\frac{\dd\sigma}{\dd\Omega}(pp\to
\pi^+d)\:.\label{equ:d_pn}
\end{eqnarray}
Here $x$ denotes the excitation energy $Q$ in the $np$ system in units of the
deuteron binding energy $B_t$, $x=Q/B_t$, and $p_{\pi}(x)$ and $p_{\pi}(-1)$
are the pion cm momenta for the $pn$ continuum and deuteron, respectively. In
a single-channel situation, the normalization $F(x$=--1$)=1$ at the deuteron
pole but it was argued~\cite{FAL1997} that deviations from this value should
be small at low $x$ if the pion production operator is of short range and the
tensor force linking the $S$ and $D$ states in the deuteron could be
neglected. However, although the shape of the high resolution inclusive data
at 951~MeV, where only the $\pi^+$ was measured~\cite{ABD2005}, is plausibly
described by Eq.~(\ref{equ:d_pn}) up to an excitation energy of $Q\approx
20$~MeV, reproducing the average absolute magnitude for $x\lesssim 9$
requires $F(x)=2.2\pm 0.1$. It should here be noted that the contribution
from $S$-wave spin-singlet $np$ pairs was shown to be very small at
951~MeV~\cite{ABD2005}.

Using bubble chamber data on single pion production in proton-proton
collisions at three energies between 900 to
1000~MeV~\cite{ERM2014,ERM2011,ERM2017}, it is clear that the excess of
$F(x)$ over unity for low $Q$ is primarily due to higher partial waves in the
final proton-neutron system~\cite{FAL2017}. This conclusion was based upon a
study of the angular distribution of the $pn$ relative momentum in the rest
frame of the two nucleons. It is thus important to see if extra information
could be obtained through a study of the dependence of the data on the
excitation energy in the $pn$ system.

\begin{figure}[h!]
\begin{center}
\includegraphics[width=1.0\columnwidth]{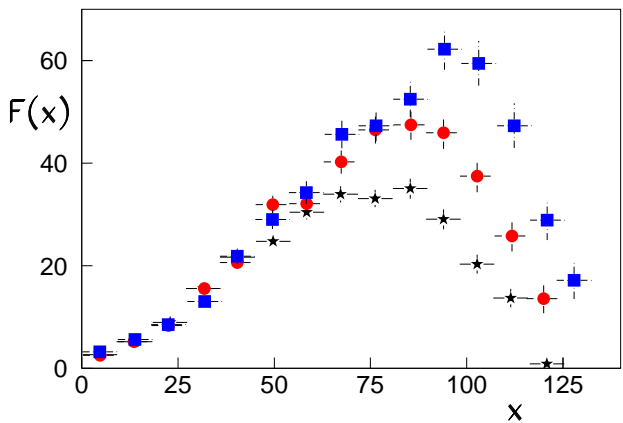}
\caption{\label{fig:raw} Values of $F(x)$ extracted from the PNPI bubble
chamber data at 900.2~MeV~\cite{ERM2014} (black stars),
940.7~MeV~\cite{ERM2011} (red circles), and 988.6~MeV~\cite{ERM2017} (blue
squares) using Eq.~(\ref{equ:d_pn}), The errors shown do not
include those arising from the numbers of $pp\to d\pi^+$ events measured
since these affect all the data at a given beam energy.}
\end{center}
\end{figure}

Figure~\ref{fig:raw} shows the values of $F(x)$ extracted from the PNPI
bubble chamber data at three beam energies~\cite{ERM2014,ERM2011,ERM2017}
using Eq.~(\ref{equ:d_pn}). Although we are mainly interested in the behavior
at small $Q$, it is clear from the figure that there remains a significant
dependence on the beam energy, though the overall errors arising from the
small numbers of measured $pp\to d\pi^+$ events have not been included. In
particular the data at large $x$ show an effect that seems to be linked to
the finite phase space. The effect is not caused by approximations in
Eq.~(\ref{equ:d_pn}) regarding the phase space limits but rather it is due to
the fact that the reaction is dominated by the excitation of the
$\Delta(1232)$, which necessarily involves a $p$-wave pion and hence a pion
momentum factor in the production amplitude. The pion momentum vanishes at
the kinematic limit of large $x$ and this feature leads to the maxima seen in
Fig.~\ref{fig:raw}. It is therefore more appropriate to consider
\begin{equation}
\label{equ:modified}
F^{\ast}(x)=\left(\frac{p_{\pi}(-1)}{p_{\pi}(x)}\right)^{\!\!2}F(x),
\end{equation}
which takes the $p$-wave nature of the pion production into account. Note,
however, that $F^{\ast}(x$=--1$)=F(x$=--1$)$ so that the extrapolation to the
deuteron pole is unaffected by the modification introduced through
Eq.~(\ref{equ:modified}).

\begin{figure}[htb]
\begin{center}
\includegraphics[width=1.0\columnwidth]{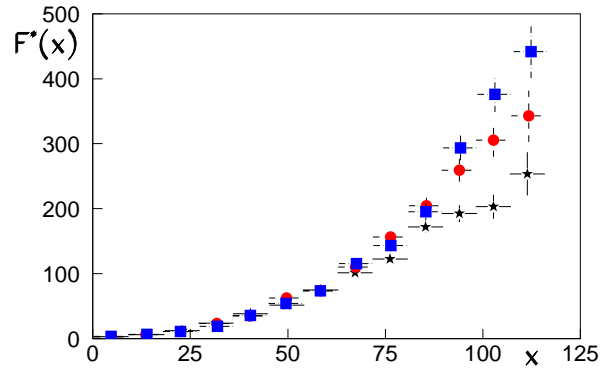}
\caption{\label{fig:mod} Values of $F^*(x)$ extracted from the PNPI bubble
chamber data at 900.2~MeV~\cite{ERM2014} (black stars),
940.7~MeV~\cite{ERM2011} (red circles), and 988.6~MeV~\cite{ERM2017} (blue
squares) on the basis of Eqs.~(\ref{equ:d_pn}) and (\ref{equ:modified}). The
errors shown do not include those arising from the numbers of $pp\to d\pi^+$
events measured.}
\end{center}
\end{figure}

The variation of $F^{\ast}(x)$ with $x$ is illustrated in Fig.~\ref{fig:mod}.
The modification introduced through Eq,~(\ref{equ:modified}) removes the
maxima seen in Fig.~\ref{fig:raw} and the data increase up to the largest
value of $x$ allowed by the kinematics. Of crucial importance is the fact
that the data at different beam energies now overlap much better so that
$F^*(x)$ is a more universal observable.

The $F^*(x)$ data of Fig.~\ref{fig:mod} were fit in the range $20<Q<180$~MeV
to the quadratic form
\begin{equation}
\label{equ:fit}
F^{\ast}(x) = A +B(x+1) +C(x+1)^2
\end{equation}
where, according to the FSI theorem~\cite{FAL1997}, the value of the
parameter $A$ should be unity. Given the wide range of $F^*(x)$ shown in the
figure, it is not surprising that this value was not confirmed by the data;
free fits give $A=3.3\pm1.6$, $9.0\pm2.0$, and $12.0\pm2.3$ at the three
energies. The error bars should be treated with some caution because the
values obtained for $A$ change significantly if higher order polynomials are
used in the fit.

\begin{figure}[htb]
\begin{center}
\includegraphics[width=1.0\columnwidth]{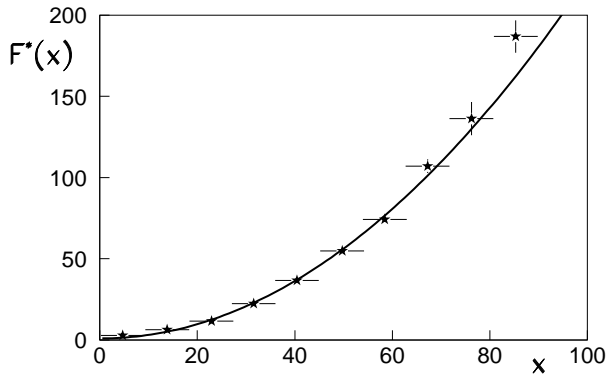}
\caption{\label{fig:sum} Values of $F^{\ast}(x)$ averaged over the PNPI beam
energies of 900.2~MeV, 940.7, and 988.6~MeV~\cite{ERM2014,ERM2011,ERM2017}
obtained using Eqs.~(\ref{equ:d_pn}) and (\ref{equ:modified}). The curves are
fits made on the basis of Eq.~(\ref{equ:fit}) with $A=1$ (fixed), $B=-0.054$,
and $C=0.0223$.}
\end{center}
\end{figure}

Imposing the constraint $A=1$ on the average of the three data sets shown in
Fig.~\ref{fig:sum} leads to a reasonable description of the data for
$20<Q<180$~MeV with $B=-0.054\pm0.026$ and $C=0.0223\pm0.0007$. The data at
larger values of $x$ clearly require a higher order polynomial to achieve an
acceptable description. Thus it is clear that the $Q>20$~MeV data are not
inconsistent with the FSI prediction of $A=1$ but, in view of the limited
statistics, they cannot provide any supporting evidence.

The principal drawback in using Eq.~(\ref{equ:d_pn}) is that it only leads to
estimates of the cross section where the final proton-neutron pair in the
$pp\to pn\pi^+$ reaction is in a relative $S$-state. A very useful tool for
investigating the effects of higher $pn$ waves was developed by Gottfried and
Jackson~\cite{GOT1964}. They defined an angle $\theta_p$, which is that
between the final proton and the incident beam direction in the $pn$ rest
frame. Any deviation from isotropy in this angle is unambiguous evidence for
the production of higher partial waves in the final $pn$ system.

\begin{figure}[htb]
\begin{center}
\includegraphics[width=1.0\columnwidth]{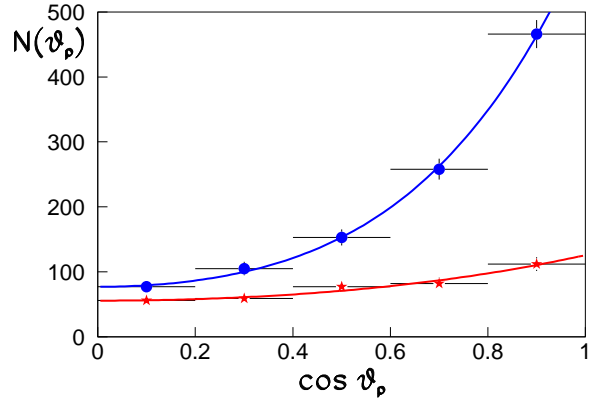}
\caption{\label{fig:zang} Distributions in the Gottfried-Jackson angle
$\theta_p$ for all the $pp\to pn\pi^+$ PNPI bubble chamber events in the
range 900--1000~MeV~\cite{ERM2014,ERM2011,ERM2017}. The (red) stars
correspond to data chosen with $0<Q<20$~MeV whereas the (blue) circles are
those where $140 < Q <160$~MeV. The lines represent the Legendre polynomial
curves of Eq.~(\ref{GJ}), with the values of the fitted parameters $C_2$ and
$C_4$ being shown in Fig.~\ref{fig:cg}.}
\end{center}
\end{figure}

It was already pointed out~\cite{FAL2017} that, even for $Q<20$~MeV, the
distributions of the bubble chamber events~\cite{ERM2014,ERM2011,ERM2017}
were not isotropic. In order to increase the statistics, in
Fig.~\ref{fig:zang} the events at all three beam energies are combined and
these show how the anisotropy grows from $Q\sim 10$~MeV to $Q\sim 150$~MeV.
It should be noted that, since the protons in the initial state are
identical, the distribution must be symmetric about $90^{\circ}$ so that the
experimental data have been folded about this point.

\begin{figure}[htb]
\begin{center}
\includegraphics[width=1.0\columnwidth]{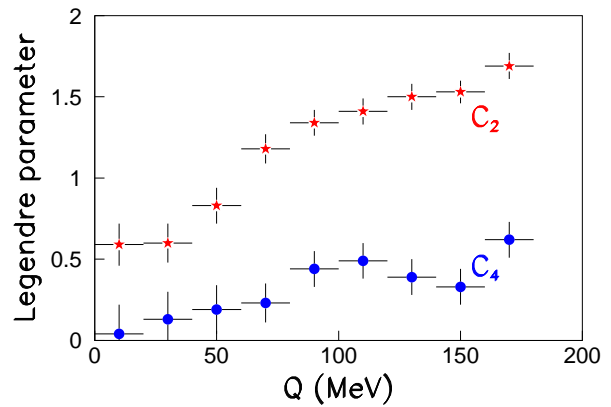}
\caption{\label{fig:cg} Variation of the parameters $C_2$ (red stars) and
$C_4$ (blue circles) of the Legendre polynomial fit of Eq.~(\ref{GJ}) to the
Gottfried-Jackson angular distribution observed in the PNPI bubble chamber
measurements of the $pp\to pn\pi^+$ differential cross
section~\cite{ERM2014,ERM2011,ERM2017}.}
\end{center}
\end{figure}

The distributions in the $\theta_p$ angle were fitted with the Legendre
polynomial expansion
\begin{equation}
\label{GJ}
N(\theta) = C_0[1 + C_2 P_2(\cos\theta) + C_4 P_4(\cos\theta)]
\end{equation}
and the values of the parameters $C_2$ and $C_4$ thus obtained are plotted 
in Fig~\ref{fig:cg} for the different energy bins. The resulting curves 
and data are also shown at two energies in Fig.~\ref{fig:zang}.

Any non-vanishing of the $C_2$ or $C_4$ parameter would be an indication of
higher partial waves in the $np$ system but, since it is impossible from
these data to distinguish the squares of $P$-waves from $S$--$D$
interference, one can at best only establish weak lower bounds on any such
contribution. On general grounds one expects that close to threshold the
$C_2$ parameter should vary like $Q$ and $C_4$ like $Q^2$. The obvious
deviation from this rule seen in Fig~\ref{fig:cg} is in the value of $C_2$ in
the lowest $Q$-bin, which shows that higher $pn$ partial waves are important
for $Q$ even below 20~MeV~\cite{FAL2017} and this might be connected with the
strong tensor force in the spin-triplet $pn$ system.

\section{Determining the neutron-proton scattering parameters from the
$\boldsymbol{pp\to pn\pi^+}$ reaction} \label{sec3}

The $Q$ dependence of the $pp\to pn\pi^+$ cross section is sensitive to the
low energy $pn$ scattering parameters and a similar sensitivity is expected
in the $pp\to \Lambda p K^+$ reaction provided that the $\Lambda p$ system
emerges in a relative $S$-wave. The angular distribution shown in
Fig.~\ref{GJ} proves that this is not a valid assumption for $Q<20$~MeV in
the $pp\to pn\pi^+$ case~\cite{FAL2017} and there are similar doubts for
$pp\to \Lambda p K^+$ when $Q<40$~MeV~\cite{HAU2017a}.

Even if we could identify the $\Lambda p$ $S$-wave contribution in say the
$pp\to \Lambda p K^+$ reaction, there are still difficulties in extracting
the $\Lambda p$ scattering length due, in part, to the finite range of
$\Lambda p$ invariant masses accessible and the coupling to the inelastic
channels, as well as to the limited mass resolution. An alternative approach
has been advocated that exploits the analyticity of the amplitudes through
the use of a dispersion relation~\cite{GAS2004,GAS2005}. It was shown that
estimates of the $S$-wave $\Lambda p$ phase shift can be deduced from the
$pp\to \Lambda p K^+$ data through
\begin{eqnarray}
\nonumber
\frac{\delta(k)}{k}&=&-\frac1{2\pi}\sqrt{\frac{m_{\rm min}}{m_{\rm red}}}\,
\int_{m_{\rm min}^2}^{m_{\rm max}^2}\dd\mu^2
\sqrt{\frac{m_{\rm max}^2-{m_X}^2}{m_{\rm max}^2-\mu^2}}\times\\
& & \hspace{-0.5cm}
\frac{1}{\sqrt{\mu^2-m_{\rm min}^2} \ (\mu^2-{m_X}^2)}
\log{\left\{\frac{A(\mu)}{A(m_X)}\right\}},
\label{final}
\end{eqnarray}
where the principal value integral in the original derivation has been here
replaced by a standard Riemann integral~\cite{FAL2016}.

In the above, $k$ is the relative momentum in the $\Lambda p$ system and
$m_{\rm red}$ is the corresponding reduced mass so that,
non-relativistically, $Q=k^2/2m_{\rm red}c^2$. $A(\mu)$ is the enhancement
factor of the $pp\to \Lambda p K^+$ cross section with respect to phase space
as a function of the $\Lambda p$ invariant mass $\mu$. If one only takes the
$\Lambda p$ final state interaction into account then $A \propto
|J(k)|^{-2}$, where $J(k)$ is the $\Lambda p$ Jost function. The lower limit
of the integration is $m_{\rm min}=m_{p}+m_{\Lambda}$ and the $\Lambda p$
mass is fixed by the external kinematics as $m_X=m_{\rm min} + Q/c^2$.
Ideally, the upper integration limit $m_{\rm max}$ should be taken to be
infinite but in general this is not practical because of the desire to retain
only the $S$-wave in the $\Lambda p$ system. The authors of
Refs.~\cite{GAS2004,GAS2005} therefore made the approximation of cutting the
integration at $Q = Q_{\rm max}$, which they chose to be $Q_{\rm max}=40$~MeV. 
In this way they could obtain estimates for the $\Lambda p$ scattering length 
and effective range, though at the expense of introducing a theoretical 
uncertainty associated with the cut-off energy.

Even if the contribution from higher partial waves, for which there is some
evidence from the Gottfried-Jackson distribution~\cite{HAU2017a}, is
discounted, it has been shown~\cite{FAL2016} that Eq.~(\ref{final}) leads to
a much better determination of the position of the virtual pole in the
$\Lambda p$ system than it does of the scattering length or effective range
individually. We now want to apply the methodology to the $pp\to pn\pi^+$
reaction.

Apart from the increased complications due to the pion multiple scatterings,
the critical difference between the $pp\to \Lambda p K^+$ and $pp\to pn\pi^+$
reactions is that there is a true bound state, the deuteron, in the $pn$
system whereas the virtual state pole in the $\Lambda p$ case is on the
second sheet. The derivation of Eq.(\ref{final}) must therefore be modified
accordingly and this can be accomplished following Eq.~(12.63) of Newton's
book~\cite{NEW1982}. Suppose that there is just one bound state at
$k=i\alpha$. In the \emph{reduced} $S$-wave Jost function the bound-state
pole is replaced by one corresponding to a virtual state, $\alpha \to
-\alpha$, by constructing
\begin{equation}
\label{definition}
J^{\rm red}(k) = J(k)\left(\frac{k+i\alpha}{k-i\alpha}\right).
\end{equation}
For real $k$, the magnitudes of the two Jost functions are identical;
\begin{equation}
|J^{\rm red}(k)| = |J(k)|.
\end{equation}
Since the phase of the Jost function is determined by the $S$-wave phase
shift $\delta$,
\begin{equation}
J(k)=|J(k)|\exp(-i\delta),
\label{phase}
\end{equation}
it follows immediately that the reduced phase shift is related to the true
one through
\begin{equation}
\delta^{\rm red}(k) = \delta(k) - i\log\left(\frac{k+i\alpha}{k-i\alpha}\right)\!\cdot
\label{reduced}
\end{equation}

For small values of $k$, $i\log\left[(k+i\alpha)/(k-i\alpha)\right]\approx
-\pi +k/\alpha-k^3/3\alpha^3$. The $-\pi$ term, which is a consequence of
Levinson's theorem when there is one bound state, does not contribute in the
evaluation of $\cot\delta$. It can therefore be neglected so that, 
effectively, for small $k$,
\begin{equation}
\label{NEW}
\delta(k) = \delta^{\rm red}(k) +2k/\alpha -2k^3/3\alpha^3 +O(k^5),
\end{equation}
where it is $\delta^{\rm red}(k)$ that is approximated by the formulae of
Ref.~\cite{GAS2004}, with a virtual rather than a bound state.

In a low energy expansion, the $S$-wave phase shift can expressed in terms of
the scattering length $a$ and effective range $r$ as
\begin{equation}
\delta(k) = -ka +a^2k^3(a/3-r/2) + O(k^5).
\end{equation}
Using this expansion in Eq.~(\ref{NEW}) for both $\delta(k)$ and $\delta^{\rm
red}(k)$, and comparing terms of order $k$ and $k^3$, shows that
\begin{eqnarray}
\nonumber
a_1 &=& a_0 - 2/\alpha,\\
a_1^{\,2}(a_1-3r_1/2) &=& a_0^{\,2}(a_0-3r_0/2) -2/\alpha^3,
\label{wbs}
\end{eqnarray}
where $a_1$ and $r_1$ are the scattering length and effective range when the
pole is a bound state and $a_0$ and $r_0$ are the corresponding parameters
when there is a virtual bound state at $k=-i\alpha$.

It must be emphasized that, apart from the change in the sign of $\alpha$,
the parameters of the true and reduced Jost functions are identical, though
this is by no means obvious when looking at the very different values of the
scattering length and effective range determined by Eq.~(\ref{wbs}).

It would, of course, be preferable to test the methodology described above on
experimental data but, as shown by the Gottfried-Jackson
distributions~\cite{GOT1964}, higher partial waves contribute in $pp\to
pn\pi^+$ at even small values of $Q$~\cite{FAL2017}. We use instead data
generated from the one-pole Jost function, where
\begin{equation}
\label{Jost}
J(k)=\frac{k-i\alpha}{k+i\beta}.
\end{equation}
This form corresponds to the Bargmann potential where the expressions for the
scattering length and effective range are, respectively,
\begin{equation}
\label{ar}
a=\frac{\alpha+\beta}{\alpha\beta}\ \textrm{and}\ r=\frac{2}{\alpha+\beta}\cdot
\end{equation}

Although the exact numbers are not crucial for the purposes of a test, the
experimental spin-triplet values of $a=5.414$~fm and
$r=1.757$~fm~\cite{DUM1983} correspond to parameters
$\alpha=0.2315~\textrm{fm}^{-1}$ and $\beta=0.9055~\textrm{fm}^{-1}$ for the
Bargmann potential.

\begin{figure}[htb]
\begin{center}
\includegraphics[width=1.0\columnwidth]{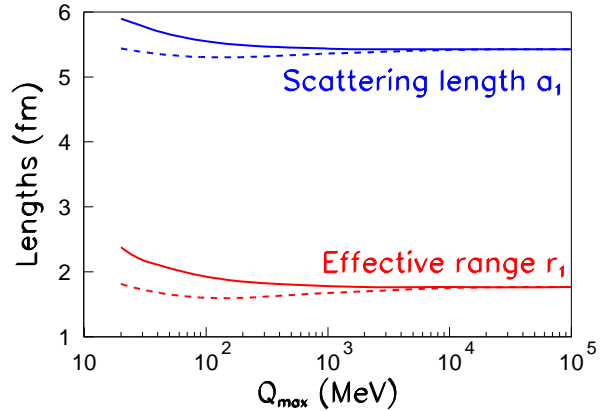}
\caption{\label{fig:pion} Estimates of the spin-triplet $np$ scattering
length ($a_1$) and effective range ($r_1$) deduced from Eqs.~(\ref{final})
and (\ref{wbs}) as functions of the cut-off energy $Q_{\rm max}$ in the
integration. The true values with no cut-off should be $a_1=5.414$~fm and
$r_1=1.757$~fm. The solid lines were obtained with the \emph{bare} Jost
function of Eq.~(\ref{Jost}) as input but with $\alpha \to -\alpha$. The
dashed lines were derived by using the modified Jost input of
Eq.~(\ref{mod_input}), where the effects of the $p$-wave nature of the pion
production have been included.}
\end{center}
\end{figure}

Estimates of $a_0$ and $r_0$ were made on the basis of the dispersion
integral of Eq.~(\ref{final}) using the Jost function of Eq.~(\ref{Jost})
with the sign of $\alpha$ changed so that there is a virtual rather than a
true bound state. The corresponding bound state case was then treated by
employing the relations given in Eq.~(\ref{wbs}). The results are shown in
Fig.~\ref{fig:pion} as functions of $Q_{\rm max}$. In order to be consistent
with the potential description, non-relativistic kinematics have been used in
the studies, even though this is hard to justify for $Q_{\rm max} \gtrsim
100$~MeV.

It must be noted that the pole position $\alpha_0$ determined from
Eq.~(\ref{final}) should be identical to that of the input because this is
determined uniquely by the singularity in $A(m_X)$, which is independent of
the value of $Q_{\rm max}$. It then follows from Eq.~(\ref{reduced}) that the
pole position of the bound state, $\alpha_1$, should be equally stable to
changes in the value of $Q_{\rm max}$. However, for small $Q_{\rm max}$,
where higher order terms in the effective range expansion become relatively
more important, it is necessary to take these into account when extracting
the value of $\alpha$.

The variation of the $np$ scattering length $a_1$ with the integration
cut-off parameter $Q_{\rm max}$ is less strong than in our previous work on
the $pp \to K^+\Lambda p$ reaction~\cite{FAL2016}. This can be linked to the
different value of $\beta/\alpha$ since Eq.~(\ref{ar}) shows that for large
$\beta$ the scattering length is fixed primarily by the value of $\alpha$,
which is very stable. These arguments do not, of course, apply to the
effective range $r_1$.

It was remarked already in the papers on the dispersion integral approach to
the analysis of $pp\to K^+\Lambda p$ data~\cite{GAS2004,GAS2005,HAU2017} that
the value obtained for the $\Lambda p$ scattering length could be distorted
through a reflection of the production of $N^*$ isobars in the $K^+\Lambda$
channel. This is typically a problem of limited energy where the Dalitz plot
is not sufficiently open and the dependence on the three invariant masses
cannot be independently determined. We saw in the description of the
bubble chamber data on $pp\to \pi^+pn$~\cite{ERM2014,ERM2011,ERM2017} in
Fig.~\ref{fig:mod} that it was important to take the $p$-wave nature of the
$\Delta(1232)$ into account and so it is interesting to study the estimates
made in the dispersion relation approach using as input an enhancement factor
of the form
\begin{equation}
\label{mod_input}
A = \left(\frac{k^2+\beta^2}{k^2+\alpha^2}\right)(p_{\pi})^2\,,
\end{equation}
where the pion momentum $p_{\pi}$ is a function of $x$ and hence of $k^2$.

The effect of the pion momentum factor depends on the limits of phase space
and, at very high energies, $p_{\pi}$ is essentially constant over the
relevant part of the integration in Eq.~(\ref{final}), in which case the
results would be indistinguishable from those obtained using the
\textit{bare} Jost function. For the purposes of the test we have assumed
that the available energy is twice that of the cut-off energy $Q_{\rm max}$.
The resulting estimates for the scattering length and effective range are
compared in Fig.~\ref{fig:pion} with the predictions of $a_1$ and $r_1$
obtained without the pion momentum factor.

Just as for the \textit{bare} Jost input, within numerical uncertainties the
position of the virtual state pole remains stable at
$\alpha_0=-0.2135$~fm$^{-1}$ and $\alpha_1=-\alpha_0$. On the other hand, if
one derives values of the parameter $\beta$ from Eq.~(\ref{ar}) it is seen
that these are identical for the virtual and true bound state cases,
\textit{i.e.}, $\beta_1=\beta_0$, and that $\beta$ approaches the input value
for large $Q_{\rm max}$. In general though the value of $\beta$ obtained in
the pion $p$-wave case is bigger than that for the \textit{bare} Jost input
because it effectively reduces the strength at larger $k^2$.

\section{Conclusions}
\label{sec4}

Using the PNPI bubble chamber data, we have investigated two features of the
$pp\to\pi^+pn$ reaction in ways that minimize the model dependence. Though we
had earlier shown that the cross section at low $np$ excitation energy $Q$
was high as compared to that of $pp\to\pi^+d$, and that this was probably
linked to the production of higher partial waves in the $np$
system~\cite{FAL2017}, studies of the cross section ratio presented here do
not clarify sufficiently the effect. However, the deviations from the
predictions of the final state interaction theorem~\cite{FAL1997,BOU1996} are
particularly large in the lowest $Q$ bin. This might arise from the strong
tensor force in the $np$ system. This discrepancy is little affected by the
distortions induced by the production of the $\Delta(1232)$ isobar.

Even though the behavior of the $pp\to\pi^+pn/pp\to\pi^+d$ ratio with $Q$
could not be investigated completely with the limited statistics of the bubble
chamber experiments, it is of interest to ask to what extent such data could
in principle be used to investigate the neutron-proton scattering length. We
generalized the dispersion relation approach of Refs.~\cite{GAS2004,GAS2005}
to situations where, as in this case, there is a true bound state in the $np$
system. The results of this, or a direct-fitting approach to the data, are
influenced by the dominantly $p$-wave nature of pion production and, if this
is not taken into account, a systematic error is made in the extracted value
of the $np$ scattering length.

Although the value obtained for the scattering length changes with the
cut-off, or whether the pion $p$-wave factor is included or not, the position
of the bound-state pole remains completely fixed, so that it is primarily
this parameter that could be fixed by the data rather than the scattering
length and effective range separately. This parallels our discussion of the
$pp\to K^+\Lambda p$ reaction, where it is the position of the virtual bound
state that could be determined by good data~\cite{FAL2016}. All this assumes,
of course, that data can be obtained with purely $S$-wave $np$ or $\Lambda p$
events. As is clear from Fig.~\ref{fig:zang}, this presents more of a
challenge as $Q$ is increased, which reinforces our argument that the data
should be used to fix the pole rather than the scattering length. This will
remain true even if the pion multiple scatterings are taken into account,
though these will undoubtedly complicate the extrapolation to the deuteron
pole.\\

\section*{ACKNOWLEDGEMENTS}
We are grateful to Dr Hauenstein for providing us with the distribution in
the Gottfried-Jackson angle for the $pp\to K^+\Lambda p$ reaction for
$Q<40$~MeV.

\end{document}